\documentclass[doublecol]{epl2} 

\usepackage{graphicx}
\usepackage{amsmath,amssymb}
\usepackage[normalem]{ulem}

\title{Optimal steering of a smart active particle}

\author{E. Schneider and H. Stark\thanks{E-mail: \email{Holger.Stark@tu-berlin.de}}}
\shortauthor{E. Schneider \and H. Stark}

\institute{                    
  \inst{1} Institute of Theoretical Physics, Technische Universit{\"a}t Berlin, Hardenbergstr. 36, D-10623 Berlin, Germany\\
}

\abstract{
We formulate the theory for steering an active particle with optimal travel time between two locations and apply it
to the Mexican hat potential without brim. For small heights the particle can cross the potential barrier, while for large heights
it has to move around it. Thermal fluctuations in the orientation strongly affect the path over the barrier. Then we consider a
smart active particle and apply reinforcement learning. We show how the active particle learns in repeating episodes to 
move optimally. The optimal steering is stored in the optimized action-value function, which is able to rectify thermal fluctuations.
}

\begin{document}

\maketitle

\section{Introduction}
Active motion, its understanding, and its phenomenology has evolved as a new paradigm in non-equilibrium physics as 
documented by recent reviews \cite{Ramaswamy2010,MarchettiSimha2013,ElgetiGompper2015,CatesTailleur2015,
ZoettlStark2016,bechinger2016active}. The control of active motion comes more and more into focus. In particular,
the motion of active particles and microswimmers is influenced by external fields, as exemplified by a few works on  
magnetic \cite{Waisbord16,Meng18}, gravitational 
\cite{Pedley92,Drescher09,Durham09,Palacci10,Enculescu11,Wolff13,TenHagen14,Ginot15,Kuhr17}, 
and flow fields\ \cite{Sokolov09,Rafai10,Zottl12,Uppaluri12,Zottl13,Lopez15,Clement16,Secchi16}. 
%
%
%
Active particles interact through chemical fields, which they produce themselves
\cite{Bocquet12,Pohl14,Saha14,Liebchen15,Simmchen16,Stark18,Stuermer19,Agudo19}.
Other articles stress the role of boundaries \cite{Ruehle18,Thutupalli18,Shen19},
the influence of a complex environment \cite{Chepizhko13,Chepizhko13a,Reichhardt14,bechinger2016active,Zeitz17}, 
and rectified motion by active ratchets \cite{DiLeonardo10,Sokolov10,Pototsky13,Kaiser14,Olson17}.

Most recent articles describe the targeted manipulation of active motion.
They actively steer Janus colloids by electric fields \cite{Mano17},
%
%
%
control the motion of Janus colloids with quorum sensing rules \cite{Baeuerle18},
design interactions by information flow between active particles \cite{Khadka18},
or describe the light-controlled assembly of active colloidal molecules \cite{Schmidt19}.
How optimal search strategies depend on environment is explored in Ref.\ \cite{Volpe17},
while Ref. \cite{Nava18} suggests
minimal navigation strategies for active particles \cite{Nava18}.
In Ref.\ \cite{Liebchen19} Liebchen and L\"owen lay the ground for optimally steering active 
particles on trajectories with shortest travel time. This gives a strong link to  optimal control 
theory \cite{BookTroeltzsch}, which has been applied to viscous flow \cite{BookSritharan},
to particle steering in inertial microfluidics \cite{Prohm13}, 
and to finance \cite{BookSoner}.


A very attractive and promising field is the application of machine or reinforcement learning to active motion. 
A comprehensive account of reinforcement learning is found in Ref.\ \cite{Sutton18}.
Self-propelled entities learn in repeating episodes to perform a prescribed task. This includes 
a glider, which learns to soar in turbulent environments \cite{Reddy16}, flow navigation of smart microswimmers 
under gravity \cite{Colabrese17}, and their navigation in a grid world \cite{Muinos18}. 


This article addresses the optimal steering of active particles in a prescribed potential landscape.
We first formulate the theory for optimizing the travel time of an active particle, the orientation of which can be controlled in 
order to steer it between two locations.
An example are magnetotactic bacteria, which align along an external magnetic field \cite{Waisbord16}.
We apply the formalism to the Mexican hat potential without brim. For small barrier heights the
active particle crosses the potential barrier on a straight path, while for large heights it has to move around it. We demonstrate how 
volatile the optimal path is to thermal fluctuations in the prescribed orientation when crossing the barrier. We then make the active particle
smart and apply reinforcement learning. We demonstrate how the active particle learns in repeating episodes to move on the optimal 
path by storing its 
knowledge in the optimized action-value function. Now, fluctuations in the prescribed orientation are rectified by
this optimized function.

%
%
%
%
%
%









%

\section{Optimal steering}
We consider an active particle that moves with swimming speed $v_0$ along an intrinsic direction given by unit vector $\bm{e}$ 
and under the influence of a potential force $\bm{F}= -\nabla U$. Its total velocity is $\bm{v} = v_0 \bm{e} + \bm{F} / \xi$,
where $\xi$ is the friction coefficient, and we neglect any thermal noise for the moment. We assume the orientation $\bm{e}$ of the particle 
can be controlled and then search for the trajectory with the optimal travel time $T$ (fastest trajectory), when the particle is 
steered from initial position $\bm{r}_i$ to final position $\bm{r}_f$,
\begin{equation}
T = \int_{\bm{r}_i}^{\bm{r}_f} dt = \int_{\bm{r}_i}^{\bm{r}_f} \frac{ds}{v} \, .
\label{eq.T}
\end{equation}

In the following we use unitless quantities by replacing $\bm{v} / v_0 \rightarrow \bm{v}$, $\bm{F} / \xi v_0 \rightarrow \bm{F}$, and
$\bm{r} / L \rightarrow \bm{r}$, where $L$ is a characteristig length. The total velocity then becomes
\begin{equation}
\bm{v} = \bm{e} + \bm{F} \, .
\label{eq.v}
\end{equation}
Parametrizing the trajectory $\bm{r}(s)$ with the arc length $s$, we can write the unit tangent along the trajectory and the 
total active-particle velocity as
\begin{equation}
\bm{t} = \frac{d \bm{r}}{ds} \quad \text{and} \quad \bm{v} = v \bm{t} \, ,
\label{eq.tang}
\end{equation}
respectively. From the square of $\bm{v}$ from Eq.\ (\ref{eq.v}) and replacing $\bm{e}$ by $v \bm{t} - \bm{F}$, we obtain 
a quadratic polynomial for the total active-particle speed $v$. We determine the zero and arrive at an
expression for the particle speed,
\begin{equation}
v = \bm{t} \cdot \bm{F} + \sqrt{1 - [\bm{F}^2 - (\bm{t} \cdot \bm{F})^2]} \, .
\label{eq.v_speed}
\end{equation}
For $\bm{F} = \bm{0}$ we correctly obtain $v=1$, while the second zero would give $v=-1$. The term in the square brackets
on the right-hand side denotes the square of the force component perpendicular to $\bm{t}$. 

Now, the variation of the travel time, $\delta T = 0$, using Eq.\ (\ref{eq.v_speed}) in Eq.\ (\ref{eq.T}) determines the fastest 
trajectory in an arbitrary potential force field. To mimimize $T$ for a two-dimensional trajectory in the $xy$ plane, we write the 
trajectory as $y=y(x)$ and discretize the travel time, where $ds = \sqrt{1+[y'(x)]^2} dx$. Starting from initial trajectories, we 
mimimize $T$ based on the routine \emph{fminunc} in the program MATLAB.

\section{Fastest trajectories} 
\subsection{Mexican hat potential without brim} \label{subsec.Mexican}
\begin{figure}
\centering
  \includegraphics[width=.85\columnwidth]{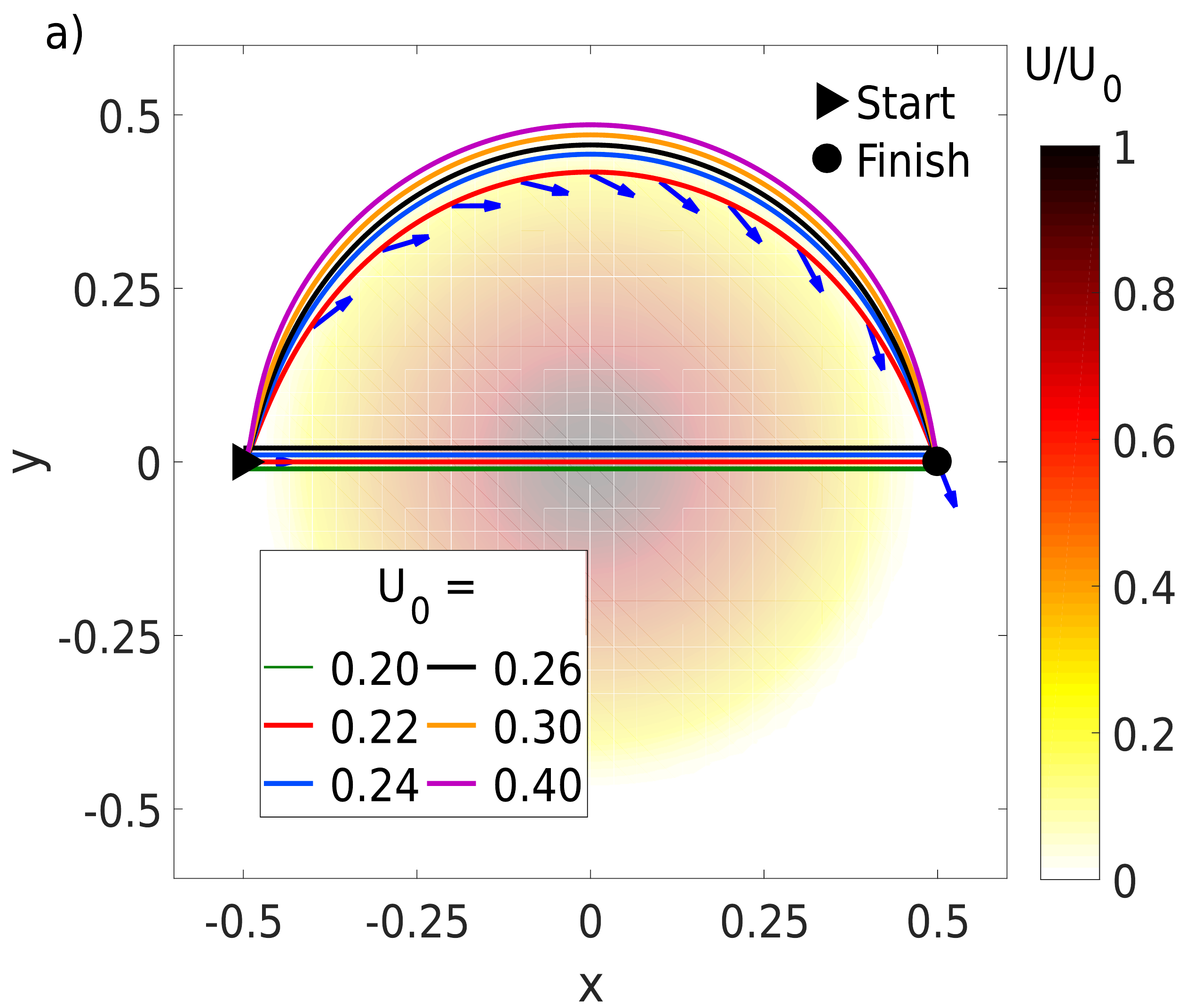}%

  \includegraphics[width=.85\columnwidth]{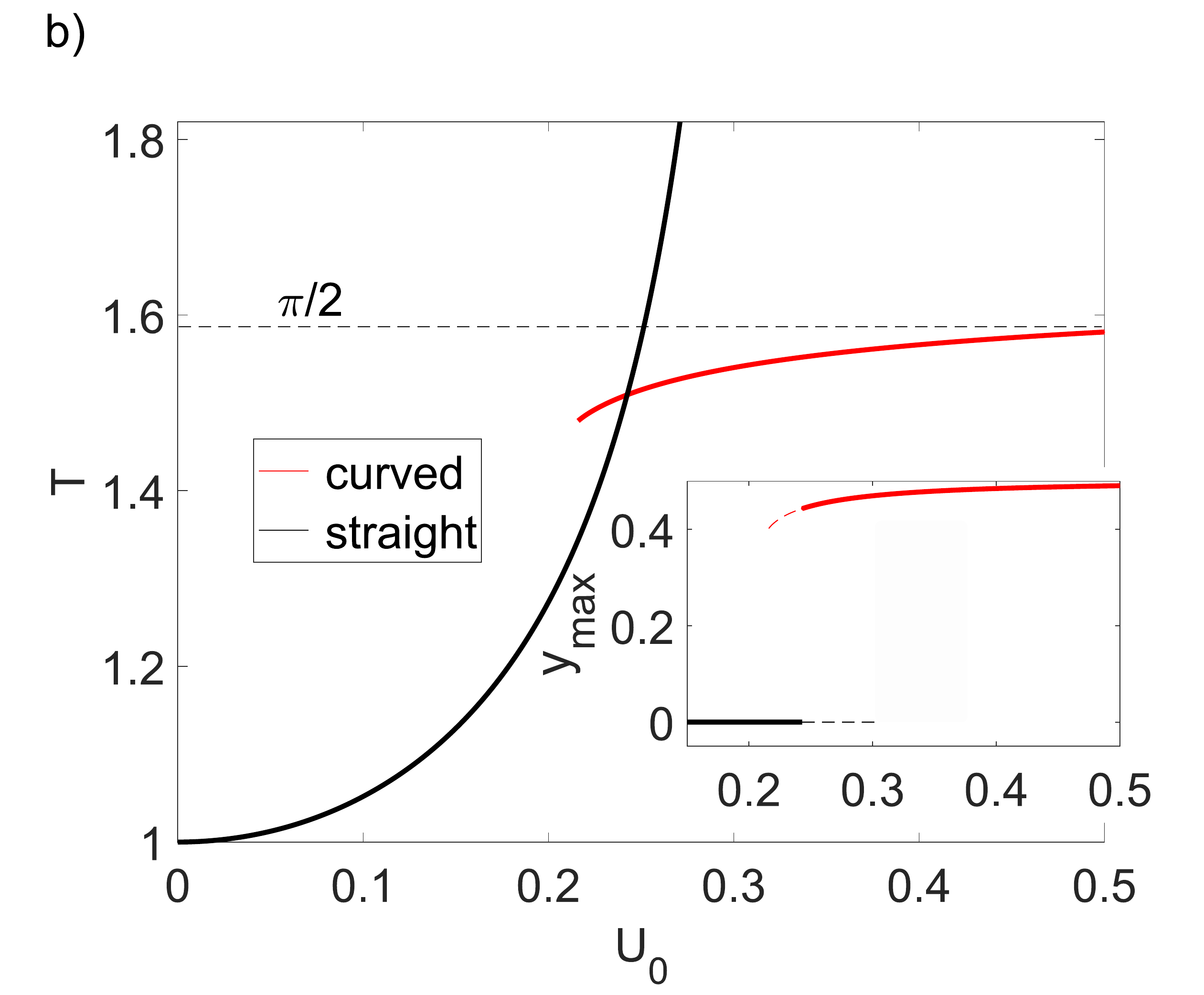}%
\caption{
a) Optimal trajectories of an active particle in a Mexican hat potential without brim (color-coded) for different $U_0$.
Trajectories with a local minimum in $T$ are shown starting at $x=-0.5$ and ending at $0.5$. For $U_0 = 0.22$ the particle
orientations $\bm{e}$ along the trajectory are indicated.
b) Travel time $T$ plotted versus $U_0$ for straight and curved trajectories. For the curved trajectory $T \rightarrow \pi /2$ for
$U_0 \rightarrow \infty$. Inset: Maximum displacement $y_\text{max}$ versus $U_0$. Solid and dashed lines mean absolute 
and metastable minima of $T$, respectively.
}
\label{fig1}
\end{figure}

To illustrate the op\-ti\-mi\-zation of the travel time, we look at active motion in two dimensions in a radially symmetric potential with
a potential barrier taken from the Mexican hat potential, while outside the minimum the potential is zero:
\begin{equation}
U = \left\{ 
\begin{array}{l}
16 U_0 (\rho^2 - 1/4)^2  \enspace , \enspace \rho \le 1/2 \\
0
\end{array}
\right.
\, ,
\label{eq.U}
\end{equation}
%
%
%
%
%
where $\rho$ is the radial distance from the center.
The potential has a barrier at $\rho=0$ with height $U_0$ and a ring of minima at 
$\rho= 1/2$.
The maximum potential force at the ring of inflection points at $\rho = 1 / 2 \sqrt{3}$ is $F_\text{max} \bm{e}_\rho = - \partial U / \partial \rho \bm{e}_\rho
=  16 U_0 / 3 \sqrt{3} \bm{e}_\rho$. Figure\ \ref{fig1}a) shows the color-coded potential. 

We now ask what is the optimal trajectory starting on the $x$ axis at $-1/2$ and ending there at $1/2$. For small height $U_0$ the 
active particle will move on a straight path. However, with increasing $U_0$ it is more favorable for the active particle to move around 
the potential barrier. Ultimately, when the active motion ($v_0$) cannot overcome the maximal drift motion $F_\text{max} / \gamma$
induced by the potential, which in our reduced units reads
$F_\text{max} = 1$, the travel time of the straight path
diverges at $U_0 = 3 \sqrt{3} / 16 \approx 0.325$. In Fig.\ \ref{fig1}a) we demonstrate optimal trajectories that evolved either from the 
straight or a curved initial trajectory for different heights $U_0$. For $U_0 = 0.22$ we also show the local orientation $\bm{e}$.
Interestingly, in the interval $U_0 \in [0.216,0.302]$ the travel times of both trajectories are local minima. In the inset of Fig.\ \ref{fig1}b)
we plot the maximal displacement in $y$ direction, $y_\text{max}$, for the local minima. Solid and dashed lines mean
absolute and metastable minima, respectively. From the main plot in Fig.\ \ref{fig1}b) the stable minima become clear. The curved 
trajectory is the fastests trajectory starting from $U_0= 0.24$ and approaches $T= \pi / 2$ for $U_0 \rightarrow \infty$, where
the active particle moves in a half circle with radius $\rho = 1/2$ around the barrier.
%
%
%
%


\subsection{More complex potential}
\begin{figure}
\centering
  \includegraphics[width=.75\columnwidth]{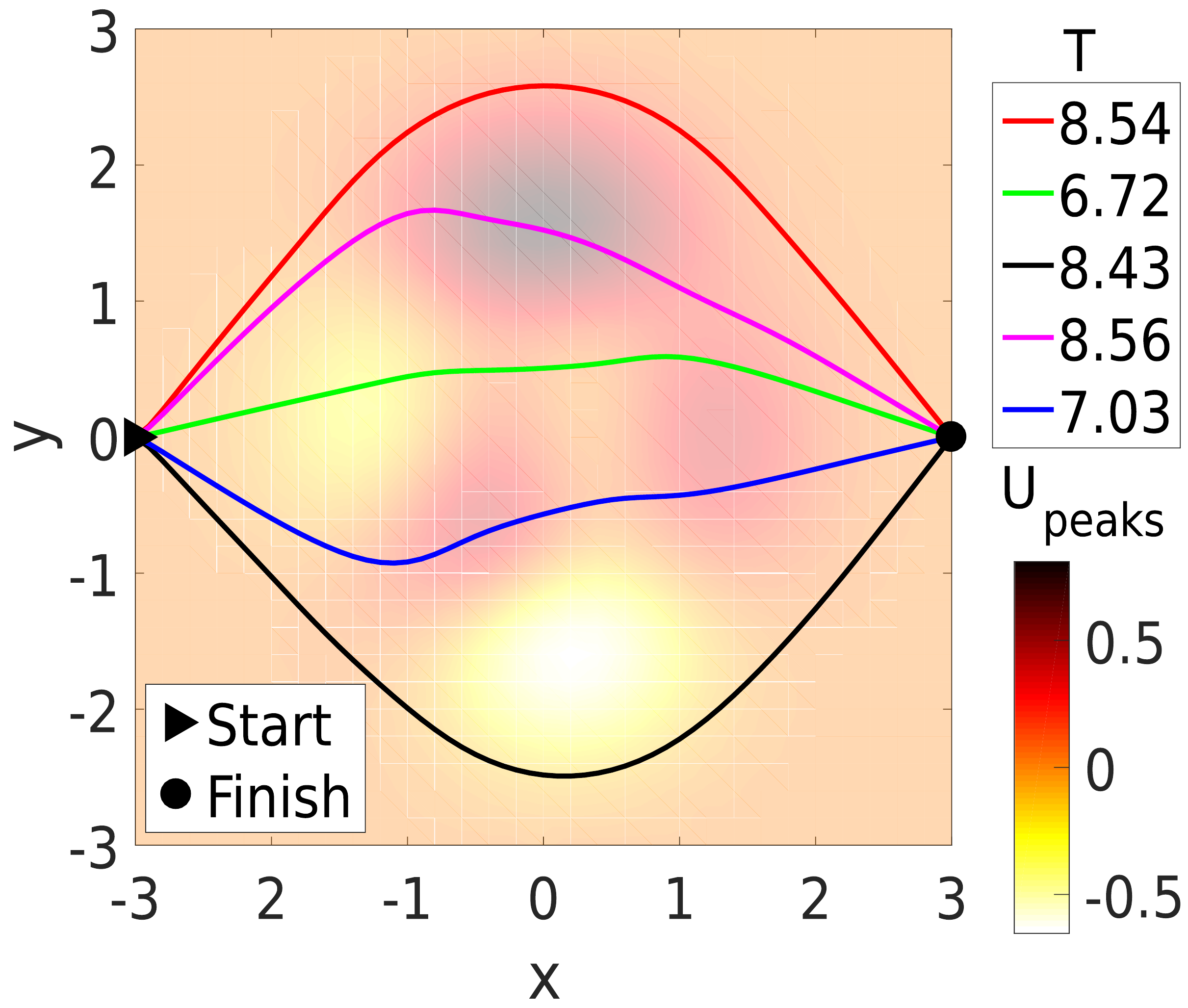}%
\caption{Locally optimal trajectories in the landscape of the peaks potential $U_\text{peaks}$ of MATLAB. The trajectories are 
obtained from different starting trajectories. The potential is shown color-coded.
%
}
\label{fig2}
\end{figure}

Of course, our method also provides optimal paths in more complex potential landscapes.
As an example we take the peaks potential provided by MATLAB,
\begin{eqnarray}
U_\text{peaks} & = & 0.3 (1- x)^2  \exp[-x^2 - (y + 1)^2] \nonumber \\ 
& & - (0.2x- x^3 - y^5) \exp(-x^2 - y^2) \nonumber \\
& &- 1/30  \exp[-(x + 1)^2 - y^2] \, ,
\end{eqnarray}
which we show color-coded in Fig.\ \ref{fig2}. The different trajectories with locally minimum travel time
evolved from different starting trajectories. Choosing them properly is crucial for finding the globally optimal
trajectory. Here, reinforcement learning can help since it provides a means to find this trajectory.


\subsection{Influence of fluctuations}
Optimal steering is hindered by noise, for example, of thermal origin. To be concrete, we ask what influence noise has on the optimal trajectories determined in the Mexican hat potential.
As mentioned in the introduction, we assume that the 
orientation of the active particle is controlled by an external field $\bm{b}$, which acts with a torque $\bm{e} \times \bm{b}$
on the orientation. Furthermore, we consider the velocity of the particle to be small enough so that the orientation adjusts quasi instantaneously to the external field along the trajectory. The optimal trajctory is then encoded in the time-dependent
field direction $\bm{b}(t) = \alpha \bm{e}_\text{opt}(t)$. We work at large P\'eclet numbers and therefore can neglect translational 
noise while rotational noise acts on the active-particle orientation, which becomes $\bm{e}(t) = \bm{e}_\text{opt}(t) + \delta \bm{e}(t)$. 
We introduce the angle $\psi$ via $|\psi | = |\delta \bm{e}|$ to quantify the orientational fluctuations relative  to $\bm{e}_\text{opt}(t)$, 
which obeys the Langevin equation
\begin{equation}
\frac{d}{dt} \psi = - \alpha \psi + \sqrt{D_R L /v_0} \, \eta \, ,
\label{eq.psi}
\end{equation}
already written in unitless quantities. Here, $\alpha$ gives the strength of the aligning torque and $D_R = k_BT/\xi_R$ is the
rotational diffusivity with $k_BT$ the thermal energy and $\xi_R$ the rotational friction coefficient. Finally, $\eta(t)$ is a Gaussian
white noise variable with unit strength. Thus, its mean vanishes, $\langle \eta \rangle = 0$, and its time-correlation function obeys
$\langle \eta(t) \eta(t') \rangle = \delta(t-t')$. For an aligning magnetic field $\bm{B}$ we have $\alpha = m B/\xi_R \cdot L/v_0$, where 
$m$ is the magnitude of the magnetic dipole moment oriented along $\bm{e}$. Typical values from Ref.\ \cite{Waisbord16} are
$m = 10^{-16} \mathrm{A m^2}$ and fields up to $\mathrm{mT}$.
%
%

\begin{figure}
\centering
  \includegraphics[width=.5\columnwidth]{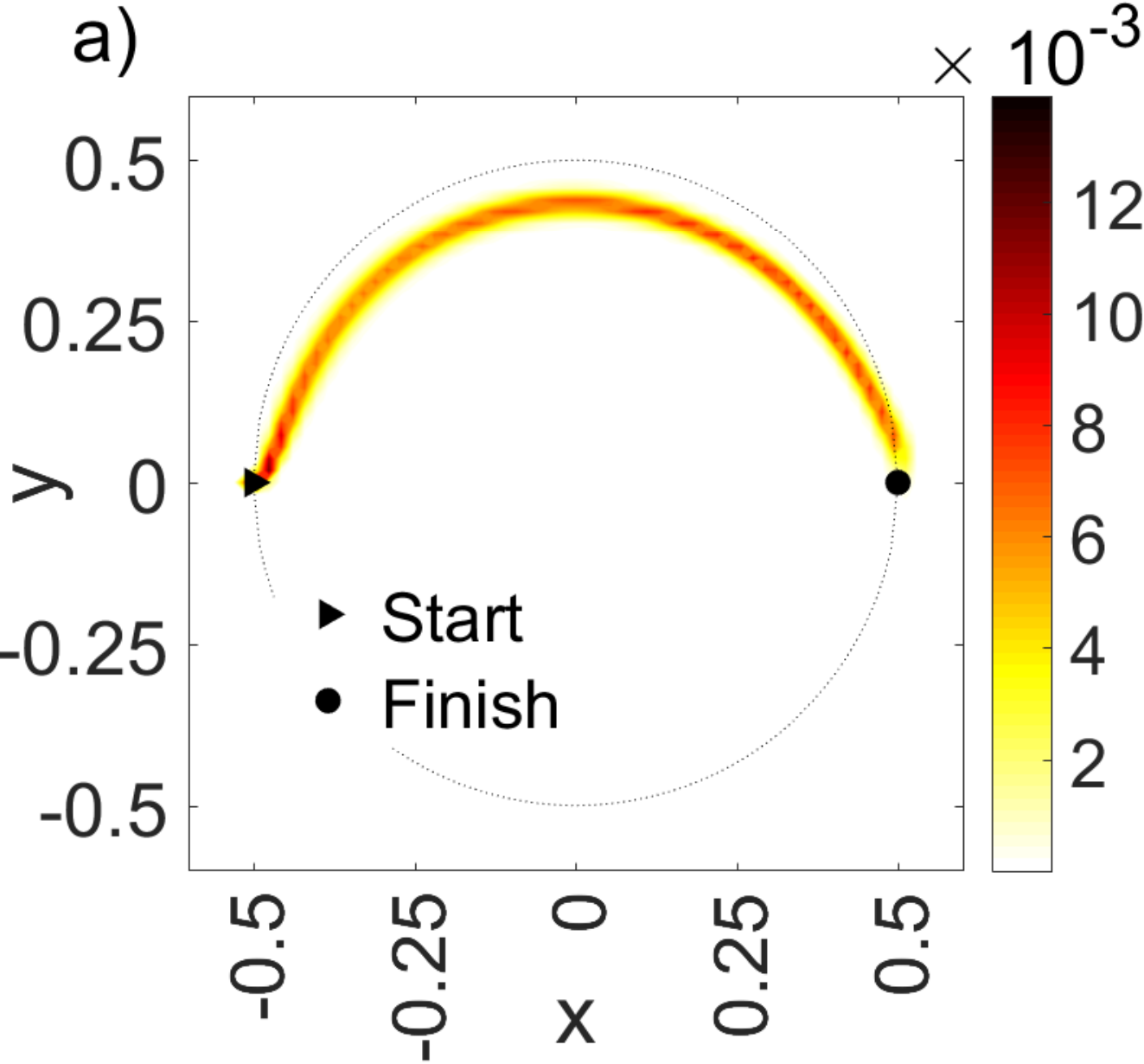}%
  \includegraphics[width=.5\columnwidth]{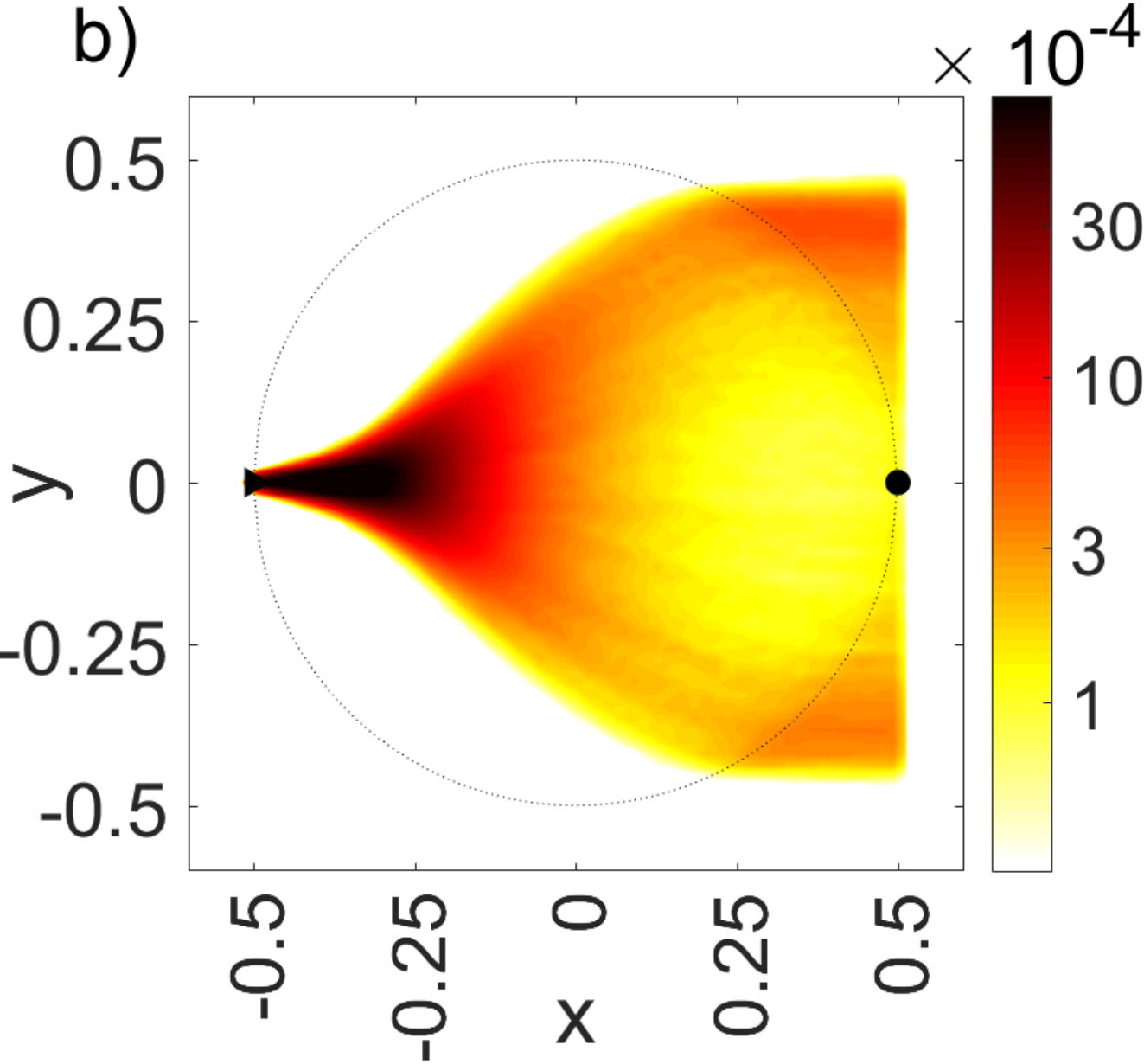}%
\caption{Heatmaps of the probability density $P(\bm{r},t)$ of the active particle for being at a location $(x,y)$ at time $t$:
a) curved and b) straight optimal paths.
The dashed line indicates the extension of the potential.
Parameters are $U_0 =0.22$, $D_R L/v_0 = 1$, and $\alpha = mB / k_BT = 10$, where 
$k_BT = 4 \cdot 10^{-21} \text{kg m}^2 / \text{s}^2 $ is thermal energy at room temperature.
}
\label{fig3}
\end{figure}

We have integrated the particle dynamics using Eq.\ (\ref{eq.v}) and the orientation unit vector $\bm{e}(t) = \bm{e}_\text{opt}(t) + 
\delta \bm{e}(t)$. Without thermal noise, $\bm{e}(t) = \bm{e}_\text{opt}(t)$, the active particle moves on the optimal trajectory.
With thermal noise included a single active-particle trajectory will deviate from the optimal path. Figure \ref{fig3} 
shows heat maps of the probability density $P(\bm{r},t)$ of the active particle for being at a location $(x,y)$ at time $t$ 
determined from 10000 simulated trajectories. The potential strength is $U_0=0.22$, where the straight trajectory is still the fastest one 
but also the curved trajectory gives a local minimum of the travel time. The aligning field strength is quantified by $mB = 10 k_\text{B}T$.
%
%
While the trajectories curving around the potential barrier are close to the optimal path (a), the straight trajectory is strongly 
disturbed by the fluctuations in the orientation and the desired final position around $(x,y) = (0.5,0)$ is hardly reached (b).
Thus, potential maxima drive active particles strongly away from optimal paths since the active particle cannot react appropriately
on the fluctuating orientation. 

Here, reinforcement learning comes in. In addition to being active the particle is also smart. It can sense its environment to 
determine its state, which in our implementation means the current location. Then the particle reacts by choosing a favorable 
orientation (action) in order to ultimately move on the optimal path. 



\section{Reinforcement Learning}
Reinforcement learning is part of the developments in machine learning and artificial intelligence \cite{Sutton18}.
It provides a means how the smart active particle (agent) learns the optimal action for each state in order to achieve
its specific goal of the fastest trajectory. To do so, the agent follows a policy, which is improved in repeated cycles of actions 
called episodes, and ultimately finds the optimal policy. In our case, an episode means moving from start to finish and
the optimal policy guides the active particle on the shortest  path.

We rely here on the method of $Q$ learning, one of the many variants of reinforcement learning \cite{Sutton18}.
The action-value function $Q(s,a)$ is a matrix, which depends on the possible states $s$ and
the possible actions $a$. In our case they mean location of the smart particle on a grid and moving to one of the 4 neighboring
grid points, respectively. Part of the policy is that for a given state $s$ the action $a$ with the largest value $Q$ is chosen,
which brings the system into a new state $s'$. Then, the action-value function  for the specific state-action pair $(s,a)$ 
is updated according to \cite{Sutton18}
\begin{equation}
Q(s,a) \leftarrow Q(s,a) + \alpha [R + \gamma \, \underset{a'}{\mathrm{max}} \, Q(s',a') - Q(s,a) ] \, .
\label{b08_eq.1}
\end{equation}
Here, $R(s,a)$ is the reward associated with performing action $a$ on state $s$ and the second term in square brackets 
takes into account the future action value associated with the new state $s'$ weigthed by the discount factor $\gamma$. 
Thus, $Q(s,a)$ increases when reward and future actions are favorable and $\alpha$ determines the learning speed. 
Starting with a constant action-value function \cite{Sutton18}, \emph{e.g.}, $Q(s,a) = 100$ at the beginning of the first episode, 
$Q(s,a)$ changes during each episode, where one tries 
to steer the active particle. Thus, the next episode starts with an ``improved'' $Q$ function. It can be proven
that $Q(s,a)$ becomes optimal after going through many episodes \cite{Jaakkola94}, 
meaning the active particle has ultimately learned to move on the optimal or shortest trajectory.

This deterministic procedure is often combined with the $\epsilon$-greedy method. In each state the action with the largest 
$Q$ value is only taken with probability $1-\epsilon$, so that also random actions are allowed. This guarantees a balance 
between exploiting the largest immediate reward but also exploring other actions, which might ultimately lead to a more optimal 
way of achieving the goal. Typically, $\epsilon$ is decreased with each episod, for example according to 
$\epsilon = 1 - i / i_\text{max}$, assuming that the $Q$ function has found its optimal value after $i_\text{max}$ episodes.



\section{Optimization by smart active particle}
\begin{figure}
\centering
\includegraphics[width=.78\columnwidth]{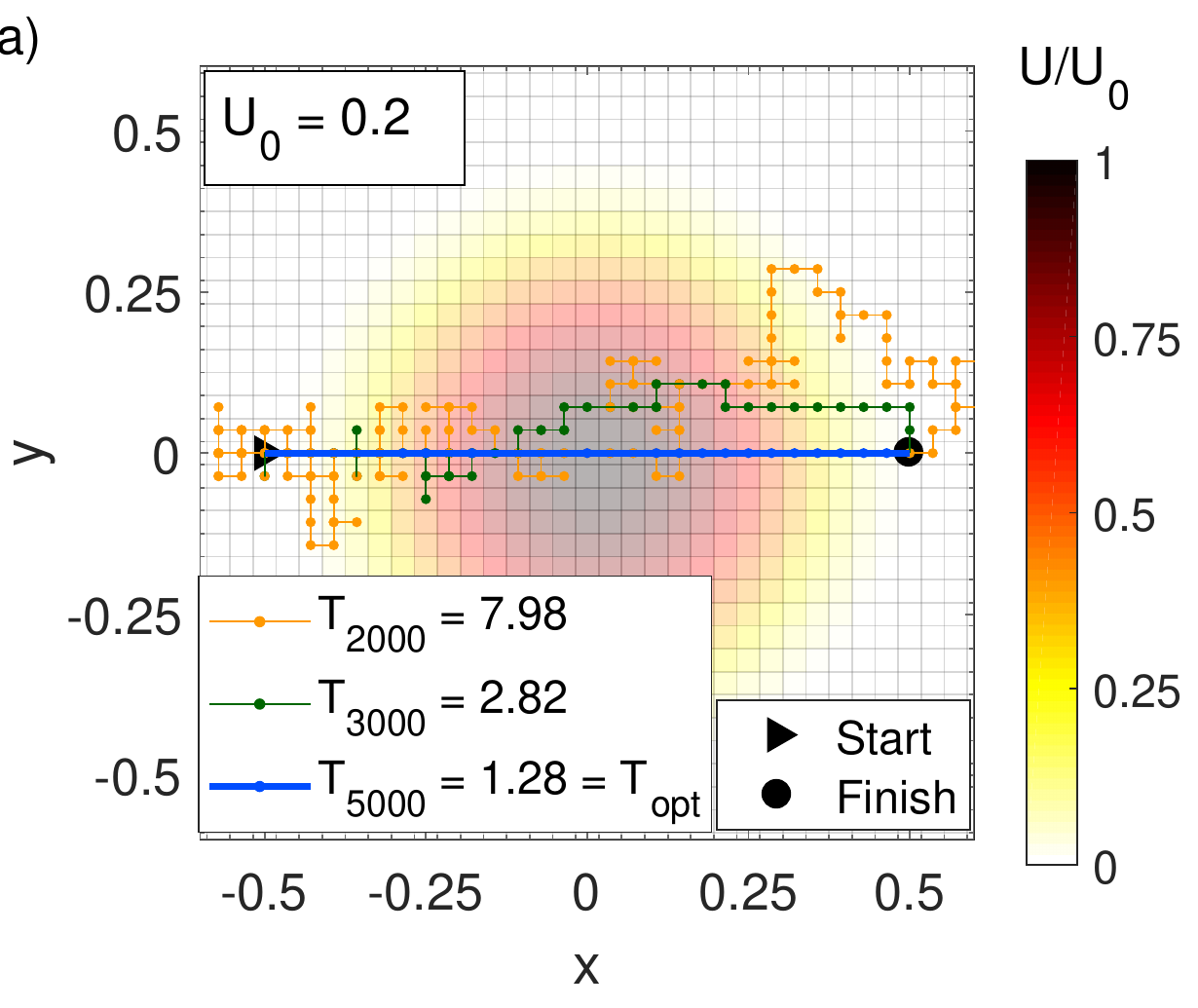}%

\includegraphics[width=.78\columnwidth]{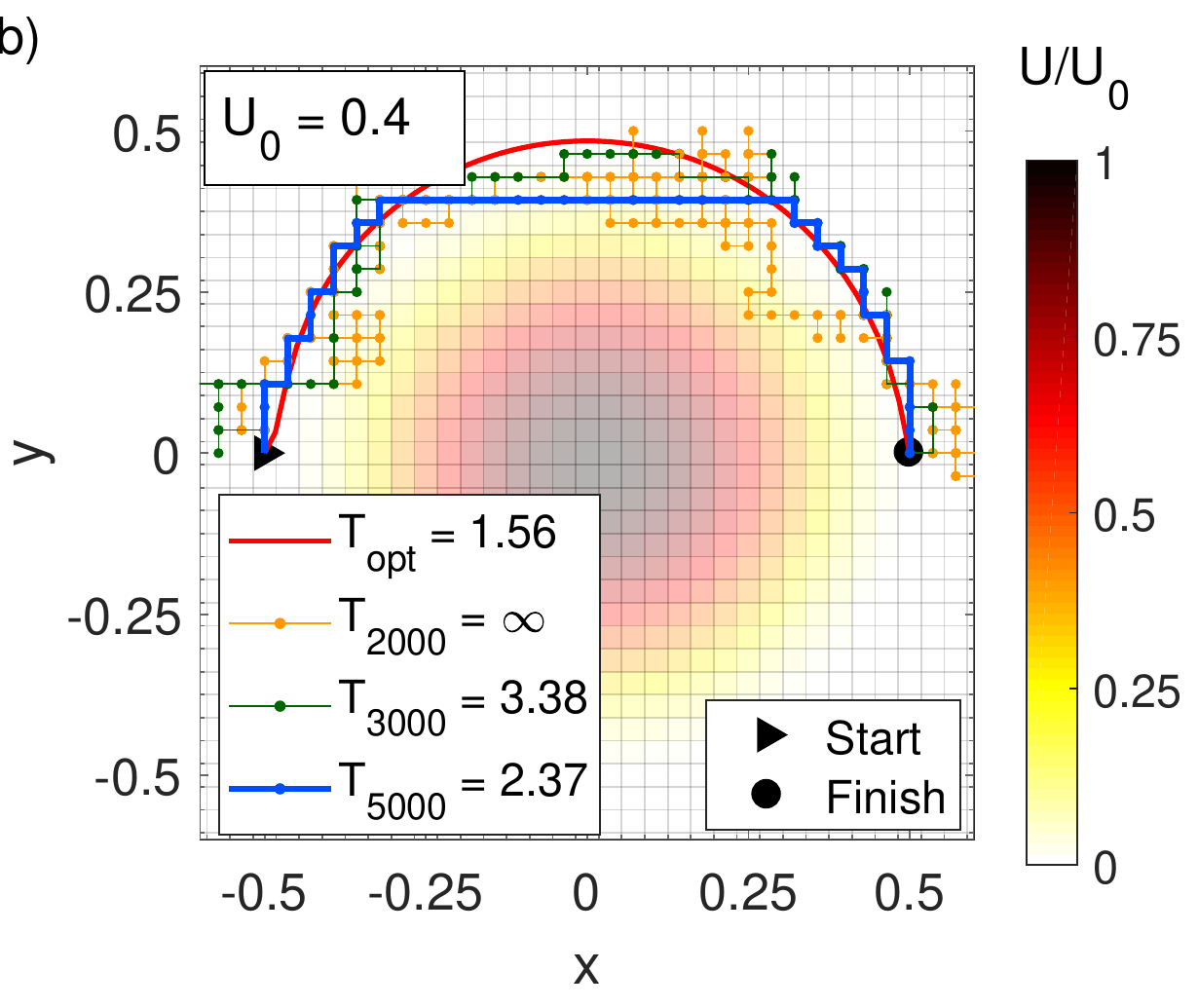}%
\caption{Trajectories of a smart active particle 
at the end of different episodes $i$ 
during $Q$ learning. The particle
moves on a grid in the Mexican hat potential without brim (color-coded) from $x=-0.5$ to $x=0.5$: a) potential strength $U_0 = 0.2$ and 
b) $U_0 = 0.4$. $T_{i}$ is the travel time of the trajectory in episode $i$ and $T_\text{opt}$ the optimal travel time. 
The travel time $T_{2000}$ in b) is infinite since the particle performs unphysical steps.
Note, the particle moves on the centers of the square unit cells.}
%
\label{fig4}
\end{figure}

We now apply the method of reinforcement learning to the optimization of the travel time of active particles in the 
Mexican hat potential of Eq.\ (\ref{eq.U}), which we treated in the beginning. We constrain the particles to move on a grid, 
which covers the plane in the region $[-0.75, 0.75] \times [-0.75, 0.75]$ and has a total of $43 \times 43 = 1849$ grid points 
with a linear spacing $a = 1.5/42 = 0.0357$ [see Fig.\ \ref{fig4}a) and b)].
The active particle can only move into four directions
to the nearest neighbours (left, right, up, and down) as dictated by the action-value function $Q(s,a)$. 
We indicate the directions by unit vectors $\bm{d}_i$.
When the particle reaches the border of the allowed region, the move out of the region is forbidden.
The particle trajectory ends when the final position is reached.

Choosing the parameters in $Q(s,a)$ is not obvious and one performs a lot of testing. Crucial is the immediate 
reward for the current action. Since we want to minimize travel time, we set $R = -dt / 1000$, where $dt = a / \bm{v} \cdot \bm{d}_i$ 
is the time to make the move to the neighboring grid point in the direction $\bm{d}_i$ under the influence of the potential force.
When calculating the velocity, we always choose the active orientation $\bm{e}$ such that $\bm{v} = \bm{e} + \bm{F} \,  \| \, \bm{d}_i$.
Clearly, if the move is fast, $R$ is only mildly negative, whereas slow moves are unfavorable. 
Note, $dt < 0$ corresponds 
to an unphysical move since it cannot be performed into the $\bm{d}_i$ direction. Thus, we choose 
a large negative reward $R = -10$ but still continue towards the end of the episod and set the travel time to infinity.  
Finally, reaching the end point is rewarded by $R=10$. Our studies with a constant greedy parameter $\epsilon = 0.6$ show 
that the learning rate $\alpha$ speeds up the convergence of the $Q$ function in time, so we choose $\alpha=0.9$ 
(see supplemental material). In more complex potentials smaller values might be useful for not getting trapped in local mimima. Furthermore, in our problem a favorable discount factor is $\gamma = 0.8$. Making it too small means that future 
rewards are not important, while choosing it close to one the immediate reward $R$ becomes less important (see supplemental
material). 
Finally, for the linearly decreasing $\epsilon = 1 - i / i_\text{max}$ in the $\epsilon$-greedy method, $i_\text{max} = 5000$ 
was chosen to be sure that the $Q$ function converged (see supplemental material). 
%

\begin{figure}
\centering
\includegraphics[width=.5\columnwidth]{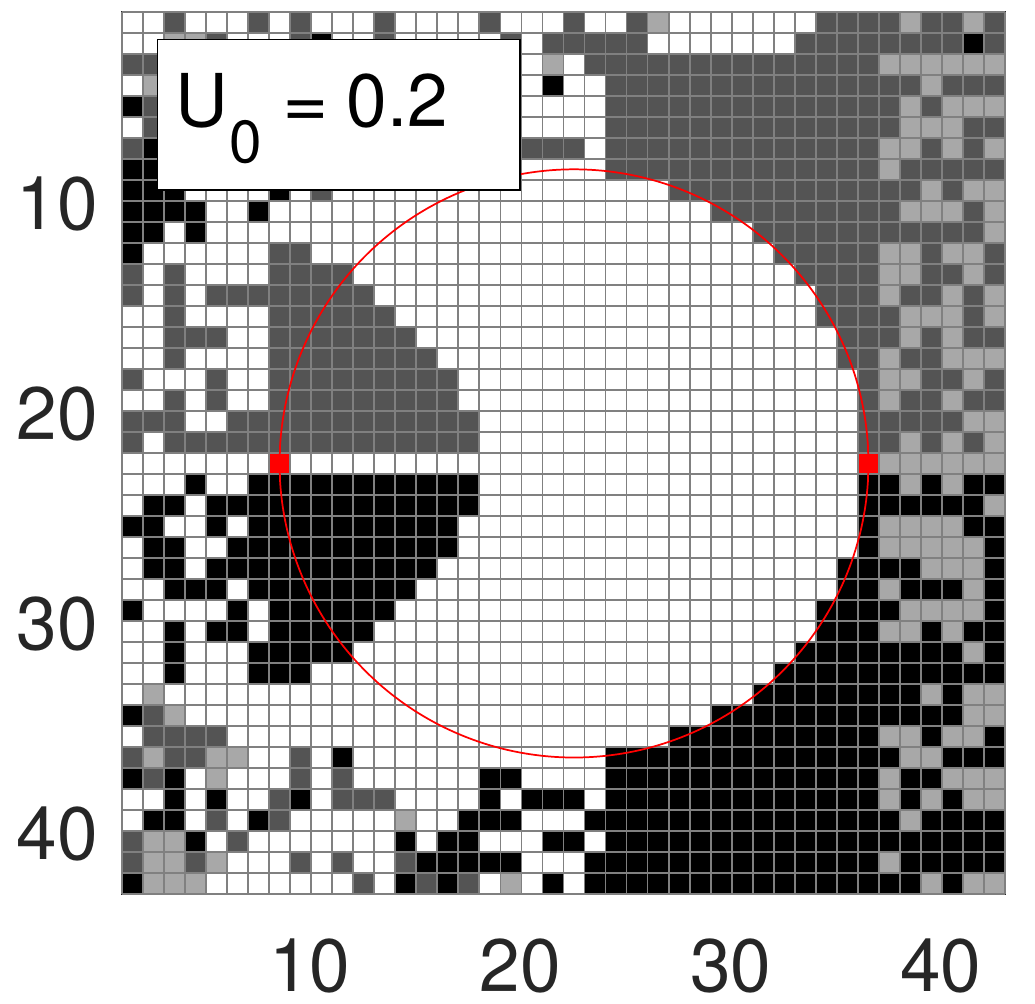}%
\includegraphics[width=.5\columnwidth]{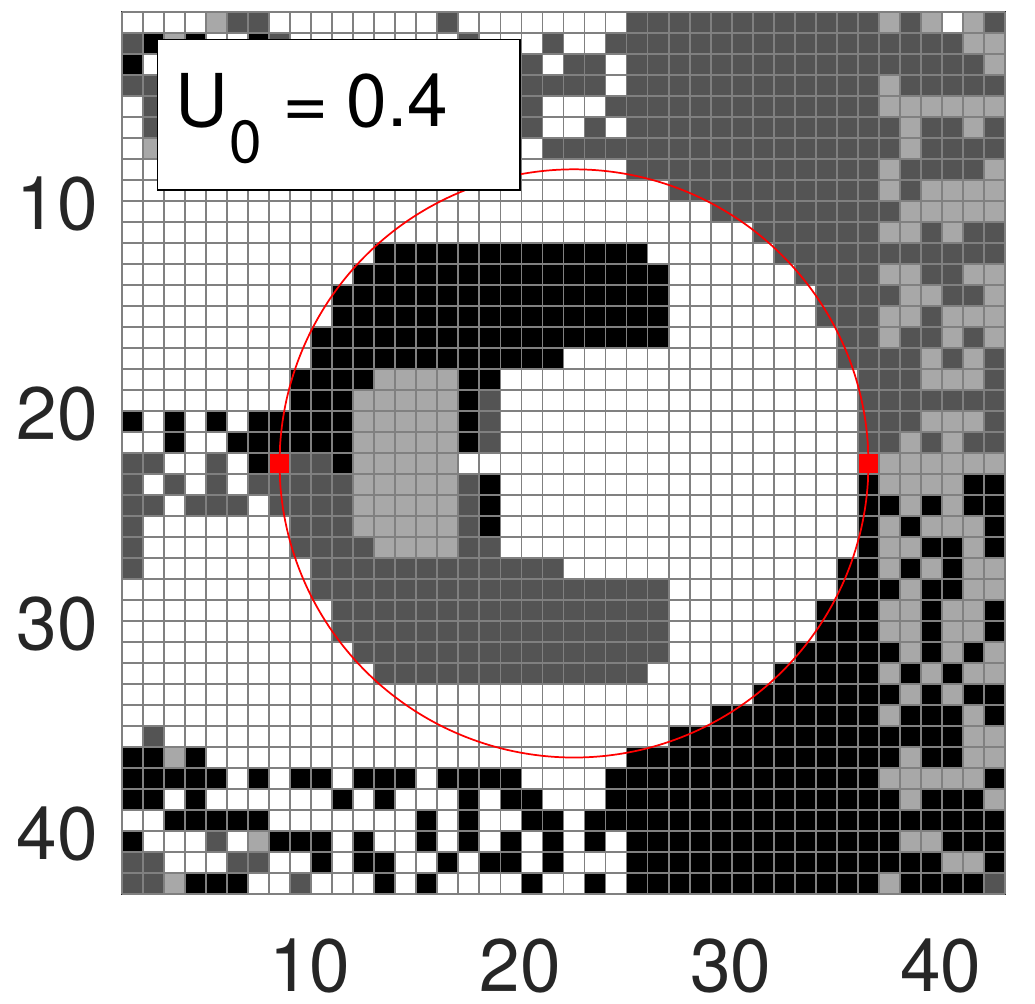}%

\vspace*{.2cm}

\includegraphics[width=.9\columnwidth]{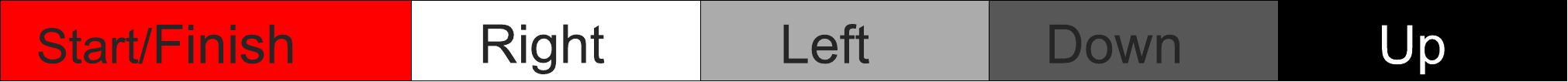}
\caption{Represenation of the optimal action-value function $Q(s,a)$ at the end of the last episode for the smart active particle 
moving in the Mexican hat potential without brim. Left: for the straight trajectory ($U_0 = 0.2$) and Right: for the curved trajectory 
($U_0 = 0.4$). For each position $s$ the action $a$ with the largest $Q$ is indicated. The matrix elements indexed by pairs 
from $[1,43] \times [1,43]$ correspond to grid points in the region $[-0.75, 0.75] \times [-0.75, 0.75]$ in the $x$-$y$ plane.
}
\label{fig5}
\end{figure}

In Fig.\ \ref{fig4} we now show two examples how the trajectories of the smart active particle evolves towards the optimal or shortest
path with increasing number of episodes. The smart active particle ``learns'' to move on the optimal trajectory and stores its
knowledge in the action-value function. In the beginning of the first episode the particle performs a random walk in the potential
since with the initial value $Q(s,a)=100$ for all state-action pairs no experience is stored (see supplemental material).
However, after going through more and more episodes the motion of the
active particle becomes directed but still contains elements of a random walk. This is still visible in Fig.\ \ref{fig4}a) for the
orange trajectory of epsiode $i=2000$. 
Ultimately, after $5000$ epsiodes the active particle has found the straight optimal path as expected for the potential height 
$U_0 = 0.2$ and the travel time coincides with the optimal value. For larger potential heights [$U_0 = 0.4$ in Fig.\ \ref{fig4}b)] we can 
also identify the curved trajectory as the optimal one. However, the optimal trajctory learned in the repeating epsiodes deviates 
from the exact solution since we only allow steps in the main directions of the grid. Adding also steps along the diagonal and
making the step size smaller would improve the shape of the optimal trajectory resulting from $Q$ learning and also better approximate 
the optimal travel time. The final or optimal action-value function $Q(s,a)$ is represented in Fig.\ \ref{fig5}. It is the result of
the learning process of the smart active particle for the optimization problem.

\begin{figure}
\centering
\includegraphics[width=.8\columnwidth]{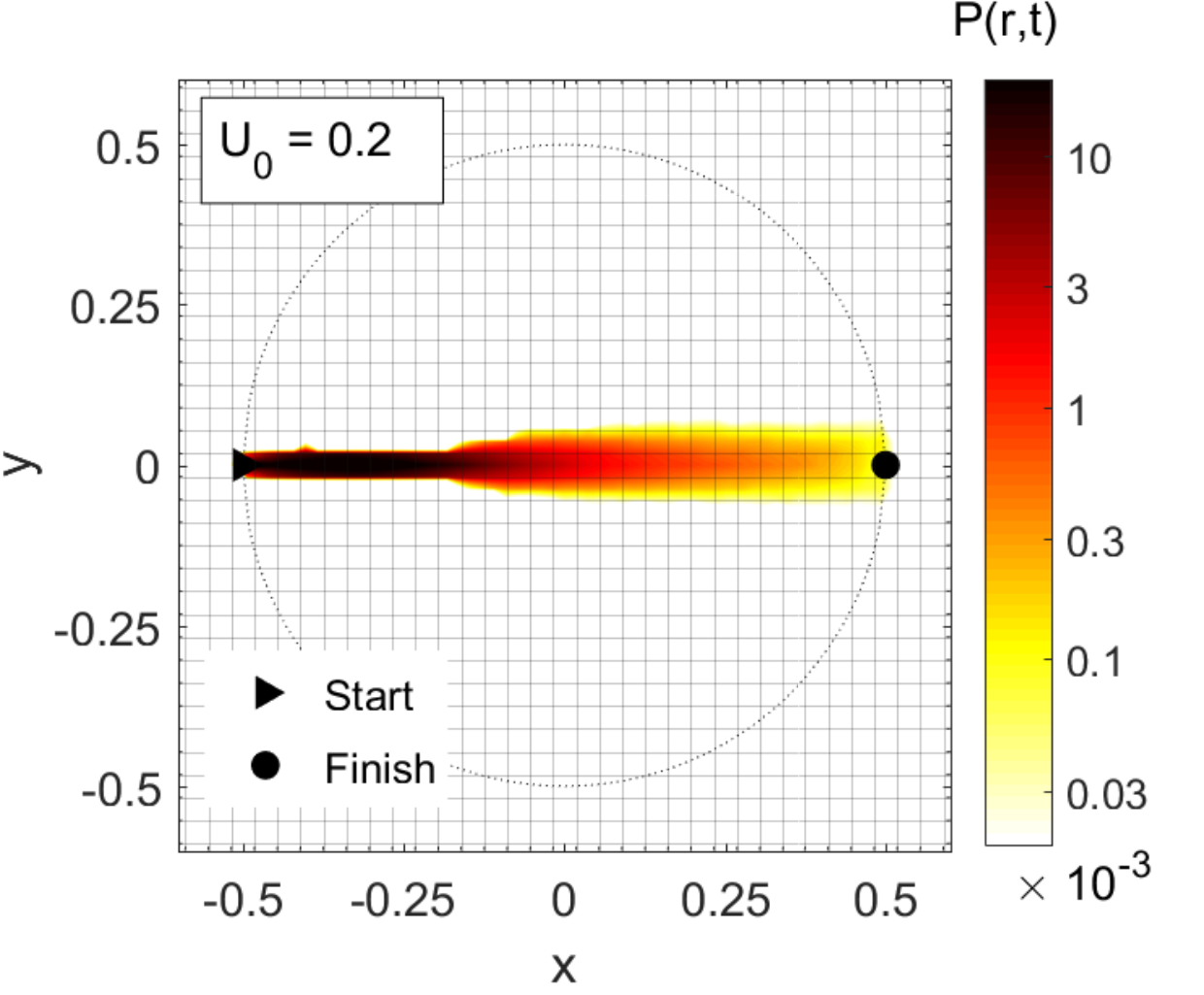}%
\caption{Heatmap of the probability density $P(\bm{r},t)$ of the active particle for being at a location $(x,y)$ at time $t$
for the straight optimal path determined by the optimal $Q$ function of Fig.\ \ref{fig5} and under the influence of thermal noise.
The dashed line indicates the extension of the potential. Parameters are $U_0 =0.2$, $D_R L/v_0 = 1$, and $\alpha = mB / k_BT = 10$, where $k_BT = 4 \cdot 10^{-21} \text{kg m}^2 / \text{s}^2 $ is thermal energy at room temperature.
}
\label{fig6}
\end{figure}

We now address the influence of (thermal) noise on the learned optimal trajectories. In Fig.\ \ref{fig3}b) we saw that the 
straight path is unstable against fluctuations. How do the optimal trajectories behave under the learned strategy in 
combination with noise? 
To investigate this, we calculate the active orientation $\bm{e}_\text{opt}$ that belongs to the 
optimal directon $\bm{d}_i$ as determined by the optimal action-value function $Q(s,a)$. However, instead of proceeding 
with the step to the center of the neighboring square unit cell,
we perform a Langevin dynamics simulation. 
We let $\bm{e}$ fluctuate around $\bm{e}_\text{opt}$ and 
integrate Eqs.\ (\ref{eq.v}) and (\ref{eq.psi}) until the particle crosses the border to one of the neighboring unit cells.
Then the process is repeated. While the heatmap for the learned curved trajectory looks similar to 
Fig.\ \ref{fig3} a),  the straight trajectory is stable under flucatuation as Fig.\ \ref{fig6} shows since the $Q(s,a)$ function brings the
smart active particle back on track when entering a neighboring unit cell.
This clearly demonstrates the advantage of $Q(s,a)$, which tells the particle how to move even on locations outside the optimal path.

In this article we formulated and discussed the optimization problem for steering an active particle on the fastest 
path in a potential landscape
and applied it to the Mexican hat potential without brim and one example of a more complicated potential. 
We demonstrated that an optimal path which crosses a potential barrier is very volatile to thermal noise added 
to the steered orientation. We then looked at a smart active particle and applied $Q$ learning as a special branch 
of reinforcement learning. We showed how in repeating episodes the 
smart particle learns to move on an optimal trajectory by storing its gained knowledge in the optimized action-value function. 
Now, thermal noise in the orientation does hardly affect the optimal path across a potential barrier since the optimized $Q$ 
function brings the smart active particle back on track.

We hope this article motivates further research on how reinforcement learning is applied to actively moving entities,
in particular, to artificial or biological microswimmers in order to navigate optimally in a complex environment. A challenge is,
of course, to develop ideas and principles to make a microswimmer smart meaning it can sense its environment and
learn the optimal action. To mimic such a smart microswimmer, systems with an external information processing as in 
Refs. \cite{Baeuerle18} and \cite{Khadka18} are a first appealing step.

\begin{acknowledgments}
We thank A. Sen for 
interesting discussions and acknowledge support from the Deutsche Forschungsgemeinschaft 
through priority program SPP 1726 (grant number STA352/11).
\end{acknowledgments}

\bibliography{literature}
\bibliographystyle{eplbib}
\

\end{document}